\documentclass[letterpaper,twocolumn]{ACESJournal}
\usepackage{color}
\usepackage{graphicx}
\usepackage{tabularx}
\usepackage[pagebackref=true]{hyperref} 
\usepackage{amsmath,bm}
\usepackage{tabto}
\title{Multi-Level Power Series Solution for Large  Surface and  Volume Electric Field Integral Equation }
\author{Y. K. Negi$^1$, N. Balakrishnan$^1$, and S. M. Rao.$^2$}
\affiliation{Supercomputer Education Research Centre \\%
Indian Institute of Science, Bangalore Karnataka, 560012 India\\%
yknegi@gmail.com, balki@iisc.ac.in}
\affiliation{Naval Research Laboratory\\%
Washington DC, 20375, USA\\%
sadasiva.rao@nrl.navy.mil}
\begin{document}
\maketitle
\begin{abstract}
In this paper, we propose a new multi-level power series solution method for solving a large surface and volume electric field integral equation-based H-Matrix. 
The proposed solution method converges in a fixed number of iterations and is solved at each level of the H-Matrix computation. The solution method 
avoids the computation of a full matrix, as it can be solved independently at each level, starting from the leaf level. Solution at each level can be used as 
the final solution, thus saving the matrix computation time for full H-Matrix. The paper shows that the leaf level matrix computation and solution with power series gives an accurate results as full H-Matrix iterative solver method. The method results in considerable time and memory savings compared to the H-Matrix iterative solver. Further, the proposed method retains the $O(NlogN)$ solution complexity.
\end{abstract}
\begin{ACESkeywords}
Method of Moments (MoM), H-Matrix, surface  electric field integral equation,volume electric field integral equation.
\end{ACESkeywords}
\section{INTRODUCTION}
With the use of ever increasing higher frequencies for various defence and civilian applications in the current world, the electrical 
size of electromagnetic scattering/radiation problem has grown drastically \cite{1, 2}. Solving the electrically large problems numerically to obtain fast and 
accurate results is the biggest challenge in the Computational Electromagnetics (CEM) community. Also, with the increase in  computing power and memory, 
the need for large-scale solution algorithms has grown even more. Out of the various numerical methods in CEM, the most popular methods are: 
a) the Finite Difference Time Domain (FDTD) \cite{3} method in the time domain and b) the Method of Moments (MoM) \cite{4} and Finite Element 
Method (FEM) \cite{5} in the frequency domain.  Traditionally, the frequency domain methods have been more popular than the time domain methods 
as most of the early experimental results were  available  in the frequency domain and validating the computational results was convenient and easy.
Out of the various frequency domain methods, MoM  based methods are highly accurate and flexible for modeling irregular structures, the MoM matrix 
can be computed with the Surface Electric Field Integral Equation (S-EFIE) for solving Perfect Electrical Conductor (PEC)  problems with surface mesh, and the
Volume Electric Field Integral Equation (V-EFIE) \cite{6} for solving  inhomogeneous dielectric   problems with volume mesh. Further, the MoM leads 
to a smaller number of unknowns compared to FEM and is free from grid dispersion error. However, the MoM matrix is a full matrix compared to a
sparse matrix for the FEM method. Hence, the solution to large size problems with MoM in electromagnetics requires  high matrix memory  
and  computation time due to the dense matrix. Note that MoM dense matrix computation, matrix vector product and storage cost scales to $O(N^2 )$ for $N$ number of unknowns. Solving the dense matrix with an iterative solver leads to $ N_{itr} O(N^2)$ calculations for $N_{itr}$ iteration with $O(N^2)$ for matrix-vector multiplication cost. With the direct solver, the complexity grows as  $O(N^3)$. Various fast solver algorithms like Multi-Level Fast Multipole Algorithm (MLFMA) \cite{6}, Adaptive Integral Method (AIM) \cite{7}, FFT \cite{8}, IE-QR \cite{9}, 
and Hierarchical Matrix (H-Matrix) \cite{10, 11, 12} have been proposed to overcome the MoM limitations of high memory and computation cost. 
Fast solver reduces the matrix memory, matrix fill time, and matrix-vector product time to $O(NlogN)$. The reduced matrix-vector product time 
improves the solution time to $ N_{itr} O(NlogN)$ for $N_{itr}$ iterations with various iterative solution methods like Bi-Conjugate Gradient 
(BiCG) or Generalized Minimum Residual (GMRES). \\

\indent Fast solvers are built on the compressibility property of the far-field interaction matrices. The compression of the far-field matrices can be done 
using analytical matrix compression methods like MLFMA or AIM, and also with numerical matrix compression methods like H-Matrix. Compared to 
analytical compression methods, numerical compression methods are easy to implement and are kernel independent. All the fast solvers depend on the 
iteration count of the iterative solution methods. The convergence of the iterations depends on the condition number of the computed MoM matrix, 
and further, for a large number of unknowns, the convergence iteration count also increases. The high iteration count can be mitigated by using various 
preconditions like ILUT, Null-Field, and Schur's complement method based preconditioners \cite{13, 14, 15}. The matrix preconditioner improves 
the condition number of the matrices and reduces the iteration count of the overall matrix solution. Despite the improvement in solution time, the use 
of preconditioners comes with the overhead of preconditioner computation time and extra preconditioner solution time for each iteration. Also, for 
the solving of a large number of unknowns, the iteration count may still be high.

Recently there has been a trend in the CEM community for the development of an iteration-free fast solver method for solving problems with a large 
number of unknowns. Various fast direct solvers \cite{16, 17} have been proposed to overcome the iteration dependency of the solution process. 
These direct solvers are based on LU decomposition and compression methods. The methods are complex to implement and give quadratic scaling 
for complex real-world problems.

In this work, we propose a Multi-Level (ML) fast matrix solution method based on the power series \cite{18,19}. The proposed method exploits the 
property of ML matrix compression of the H-Matrix. The matrix is solved for each level using the matrix computation of the leaf level only, and the 
matrix solution can be terminated at the desired level as per the required accuracy. Our experimental results show that we get good accuracy even for the
 lowest level solution. The method relies on matrix-vector multiplication at each level and using the solution of the lowest level saves matrix computation 
time and memory requirement for the overall matrix solution. \\

The rest of the paper is organized as follows. Section II  gives a summary of MoM computation for S-EFIE and V-EFIE, section III covers H-Matrix
 computation for S-EFIE and V-EFIE. The derivation of the proposed ML power series solver is given in section IV. The numerical results of the 
proposed method, and conclusion are discussed in sections V, and VI.
\section{METHOD OF MOMENTS}
MoM is a popular and efficient integral equation based method for solving various electromagnetic radiation/scattering problems. MoM can be computed using Electric Field Integral Equation (EFIE) for both surface and volume modeling. Surface modeling can be done using Rao Wilton Glisson (RWG) \cite{20} triangle basis function, whereas volume modeling can be done using Schaubert Wilton Glisson (SWG) \cite{21} tetrahedral basis function. In the case of dielectric modeling compared to S-EFIE, V-EFIE is an integral equation of the second kind and is more well-conditioned and stable. V-EFIE can model inhomogeneous bodies more efficiently than surface EFIE. In this work, we use RWG basis function for PEC surface S-EFIE modeling and SWG basis function for volume V-EFIE modeling. The surface/volume EFIE governing equation for the conductor/dielectric scattering body illuminated with the incident plane wave is given as the total electric field $(\bm{E}^{total})$ from a scattering surface/volume and is the sum of incident electric field $(\bm{E}^{inc})$ and scattered electric fields $(\bm{E}^{scatt})$.
\begin{equation}
\label{eq:1}
\bm{E}^{total}=\bm{E}^{inc}+\bm{E}^{scatt}.
\end{equation}
The scatted electric field is due to the surface current in PEC surface or volume polarization current in the dielectric media and is given as:
\begin{equation}
\label{eq:2}
\bm{E}^{scatt}=-j\omega \bm{A}(\bm{r})- \nabla \phi(\bm{r}).
\end{equation}
In the above equation $\bm{A}(\bm{r})$ is the magnetic vector potential and describes radiation of current, $\phi(\bm{r})$ is electric potential and describes associate bound charge. Applying the boundary condition for PEC structure the S-EFIE can be written as:
\begin{equation}
\label{eq:3}
\bm{E}^{inc}=j\omega \bm{A}(\bm{r})+ \nabla \phi(\bm{r}).
\end{equation}
Similarly, the V-EFIE can be written for a dielectric inhomogeneous body as:
\begin{equation}
\label{eq:4}
\bm{E}^{inc}=\frac{\bm{D}(\bm{r})}{\epsilon(\bm{r})} + j\omega \bm{A}(\bm{r}) + \nabla \phi(\bm{r}).
\end{equation}
In the above, equation $ \bm{D}(\bm{r})$ is the electric flux density and $\epsilon(\bm{r})$ is the dielectric constant of the scattering volume media. The surface current in equation (3) for PEC structure is expanded with RWG function, and similarly in equation (4) for dielectric volume structure polarization current and charge is modeled with SWG basis function. Performing Galarkin testing over each term with integrating over the surface/volume, the final system of equation boils down to the linear system of the equation as below:
\begin{equation}
\label{eq:5}
[\bm{Z}]\bm{x}=\bm{b}.
\end{equation}
In the above equation, $\bm{Z}$ is a dense MoM matrix, $\bm{b}$ is a known incident plane wave, and $\bm{x}$ is an unknown coefficient to be computed. The dense matrix leads to high cost matrix computation and memory requirement as well as solution time complexity. In the next section, we discuss the implementation of the H-Matrix for the mitigation of high cost of the conventional MoM matrix
\section{H-MATRIX}
The high cost of MoM limits its application to a few $\lambda$ problem sizes. This limitation of MoM can be overcome by incorporating fast solvers. Most of the fast solvers work on the principle of compressibility of the far-field matrices. For the implementation of a fast solver, the mesh of geometry is divided into blocks using an oct-tree or binary-tree division process and terminated at the desired level with a limiting edge or face count in each block. The non-far-field interaction blocks at the lowest level are considered near-field blocks and are in the dense matrix form. The compression of the far-field block matrix at each level can be done analytically or numerically. The system of equations in equation (5) can now be written as the sum of near-field and far-field matrix form as:
\begin{equation}
\label{eq:6}
[\bm{Z}_N+\bm{Z}_F]\bm{x}=\bm{b}.
\end{equation}
In the above equation $ \bm{Z}_N$ is a near-field block matrix and $\bm{Z}_F$ is far-field compressed block matrices for the MoM fast solver matrix. Numerical compression of far-field matrices is easy to implement and is kernel-independent. A few of the popular fast solvers using numerical compression methods are IE-QR, H-Matrix. In this work, we have implemented H-Matrix for ML matrix compression. For the ML compression computation, the mesh is divided into ML binary tree division-based subgroups. H-Matrix works on the computation of a far-field matrix for the interaction blocks satisfying the admissibility condition given in equation (7). The admissibility condition states that $\eta$ times the distance between the observation cluster ($\Omega_t$) and source cluster ($\Omega_s$)  should be greater or equal to the minimum diameter of the observation cluster or source cluster for far-field computation, where $\eta$ is the admissibility control parameter, and its value is taken as 1.0.
\begin{equation}
\label{eq:7}
\eta \; dist(\Omega_t,\Omega_s) \geq min(diam(\Omega_t),diam(\Omega_s)).
\end{equation}
The far-field matrix block compression is done in such a way that its parent interaction matrix should not be computed at the top level. Matrix compression at each level is carried out using Adaptive Cross Approximation (ACA) \cite{22} \cite{23} method. The method exploits the rank deficiency property of the far-field matrix blocks. The low-rank sub-block of the far-field $\bm{Z}_{sub}$ with $m$ rows and $n$ columns is decomposed into approximate $\bm{U}_{(m\times k)}$ and $\bm{V}_{(k\times n)}$ matrices where $k$ is the numerical rank of the low-rank sub-block far-field matrix such that $k<<min(m,n)$.  In this work, for memory savings, we only compute half of the H-Matrix \cite{12} by making the computation process symmetric, and to maintain the accuracy of the H-Matrix, we use re-compressed ACA \cite{24} for far-field block compression. The solution of the iterative solver is iteration count dependent, and further, the convergence iteration count depends on the condition number of the matrix. Also, as the number of unknowns increases, the iterating count for the convergence increases. In the next section, we discuss our proposed method, which is an iteration count and far-field level block independent solution process.
\section{MULTI-LEVEL POWER SERIES SOLUTION}
The full H-Matrix is a combination of near-field and far-field block matrices. The far-field compressed block matrices are computed for various levels, and in equation (6), the far-field matrix ($\bm{Z}_{F}$) can be further decomposed into the different matrix levels as below:
\begin{equation}
\label{eq:8}
[\bm{Z}_F]=[\bm{Z}_{F1}]+[\bm{Z}_{F2}]+[\bm{Z}_{F3}].
\end{equation}
\begin{figure}
  \begin{center}
  \includegraphics[width=5cm]{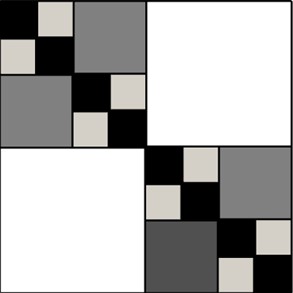}
  \caption{Compressed far-field and dense near-field matrix blocks layout.}
  \label{fig:power_reflection}
  \end{center}
\end{figure}
In the above equation far-field matrix $\bm{Z}_{F1}$ is for level 1, $\bm{Z}_{F2}$ is for level 2. and, $\bm{Z}_{F3}$ is for level 3. Level 3 forms the leaf level of the binary tree and level 1 as the top level of the tree. Fig. 1. shows the H-Matrix layout for a two-dimension strip. In Fig. 1. light gray boxes represent  $\bm{Z}_{F1}$ far-field matrix at level 1, dark gray boxes as $\bm{Z}_{F2}$ is for level 2 and large white boxes as $\bm{Z}_{F3}$ for level 3, the black boxes are the near-field dense matrices. For illustrative purposes, the near-field matrix is a diagonal block form for a two-dimension strip. The real-world problems are three-dimension in structure, giving a non-diagonal block near-field matrix. To implement our ML power series solution method, we must diagonalize the near-field block matrix. The near-field matrix in equation (6) is diagonalized using diagonal scaling coefficient $[\bm{\alpha}]$, as computed in \cite{15} such that the scaled diagonal block near-field matrix can be given as:
\begin{equation}
	\label{eq:9}
	[\tilde{\bm{Z}}_N]=[\bm{\alpha}][\bm{Z}_N].
\end{equation}
Expanding equation (8) and scaling it with the scaling coefficients $[\bm{\alpha}]$ gives:
\begin{equation}
	\label{eq:10}
	[\bm{\alpha}][\bm{Z}_{N}+\bm{Z}_{F1}+\bm{Z}_{F2}+\bm{Z}_{F3}]\bm{x}=[\bm{\alpha}]\bm{b}. 
\end{equation}
\begin{equation}
	\label{eq:11}
	[\tilde{\bm{Z}}_{N}]\bm{x}+[\bm{\alpha}][\bm{Z}_{F1}]\bm{x}+[\bm{\alpha}][\bm{Z}_{F2}]\bm{x}+[\bm{\alpha}][\bm{Z}_{F3}]\bm{x}=\tilde{\bm{b}}.
\end{equation}
In the above equation $\tilde{\bm{b}}$ is a $[\bm{\alpha}]$ scaled vector $\bm{b}	$ and can be further simplified as : 
\begin{equation}
	\label{eq:12}
	\begin{split}
	\bm{x}+ [\tilde{\bm{Z}}_{N}]^{-1}[\bm{\alpha}][\bm{Z}_{F1}]\bm{x}+[\tilde{\bm{Z}}_{N}]^{-1}[\bm{\alpha}][\bm{Z}_{F2}]\bm{x} \\
	+[\tilde{\bm{Z}}_{N}]^{-1}[\bm{\alpha}][\bm{Z}_{F3}]\bm{x}= [\tilde{\bm{Z}}_{N}]^{-1} \tilde{\bm{b}}.
\end{split}
\end{equation}
Let $[\tilde{\bm{Z}}_{N}]^{-1}[\bm{\alpha}][\bm{Z}_{F1}]=[\bm{U}_{1}]$, $[\tilde{\bm{Z}}_{N}]^{-1}[\bm{\alpha}][\bm{Z}_{F2}]=[\bm{U}_{2}]$ and $[\tilde{\bm{Z}}_{N}]^{-1}[\bm{\alpha}][\bm{Z}_{F3}]=[\bm{U}_{3}]$ equation (12) can further be simplified as
\begin{equation}
\label{eq:13}
		\bm{x}+ [\bm{U}_{1}]\bm{x}+[\bm{U}_{2}]\bm{x} +[\bm{U}_{3}]\bm{x}= [\tilde{\bm{Z}}_{N}]^{-1}\tilde{\bm{b}}.
\end{equation}
\begin{equation}
\label{eq:14}
		[\bm{I}+ \bm{U}_{1}]\bm{x}+[\bm{U}_{2}]\bm{x} +[\bm{U}_{3}]\bm{x}= [\tilde{\bm{Z}}_{N}]^{-1}\tilde{\bm{b}}.
\end{equation}
\begin{equation}
\label{eq:15}
	\begin{split}
		\bm{x}+[\bm{I}+ \bm{U}_{1}]^{-1}[\bm{U}_{2}]\bm{x} +[\bm{I}+ \bm{U}_{1}]^{-1}[\bm{U}_{3}]\bm{x}\\
		=[\bm{I}+ \bm{U}_{1}]^{-1} [\tilde{\bm{Z}}_{N}]^{-1}\tilde{\bm{b}}.
	\end{split}
\end{equation}
Let $[\bm{I}+ \bm{U}_{1}]^{-1}[\bm{U}_{2}]=[\bm{V}_{2}]$ and $[I+ \bm{U}_{1}]^{-1}[\bm{U}_{3}]\\
=[\bm{V}_{3}]$ equation (15) can further be simplified as
\begin{equation}
\label{eq:16}
	\bm{x}+ [\bm{V}_{2}]\bm{x}+[\bm{V}_{3}]\bm{x} = [I+ \bm{U}_{1}]^{-1} [\tilde{\bm{Z}}_{N}]^{-1}\tilde{\bm{b}}.
\end{equation}
\begin{equation}
\label{eq:17}
	\bm{x}+[\bm{I}+ \bm{V}_{2}]^{-1}[\bm{V}_{3}]\bm{x}=[\bm{I}+ \bm{V}_{2}]^{-1}[\bm{I}+ \bm{U}_{1}]^{-1} [\tilde{\bm{Z}}_{N}]^{-1}\tilde{\bm{b}}.
\end{equation}
Let $ [\bm{I}+\bm{V}_2 ]^{-1} [\bm{V}_3 ]=[\bm{W}_3]$ and equation (17) can be written as
\begin{equation}
\label{eq:18}
	\bm{x}+[\bm{W}_3]\bm{x}=[\bm{I}+\bm{V}_2 ]^{-1} [\bm{I}+\bm{U}_1 ]^{-1} [\tilde{\bm{Z}}_{N}]^{-1}\tilde{\bm{b}}.
\end{equation}
\begin{equation}
\label{eq:19}
		\bm{x}=[\bm{I}+\bm{W}_3 ]^{-1} [\bm{I}+\bm{V}_2 ]^{-1} [\bm{I}+\bm{U}_1 ]^{-1} [\tilde{\bm{Z}}_{N}]^{-1}\tilde{\bm{b}}.
\end{equation}
In the above equations $[\bm{I}+\bm{W}_3 ]^{-1}$,$ [\bm{I}+ \bm{V}_2 ]^{-1}$ and $[\bm{I}+ \bm{U}_1 ]^{-1}$ can be solved independently at each level using a power series solution method with the expansion as below:
\begin{equation}
\label{eq:20}
		[\bm{I}+ \bm{U}_1 ]^{-1}=[\bm{I}+ [\tilde{\bm{Z}}_{N}]^{-1}[\bm{\alpha}][\bm{Z}_{F1}]]^{-1}.
\end{equation}
\begin{equation}
\label{eq:21}
	\begin{aligned}
	[\bm{I}+\bm{V}_2 ]^{-1}=[\bm{I}+[\bm{I}+\bm{U}_1 ]^{-1} [\bm{U}_2 ]]^{-1}\\
	=[\bm{I}+[\bm{I}+ [\tilde{\bm{Z}}_{N}]^{-1}[\bm{\alpha}][\bm{Z}_{F1}]]^{-1} [\tilde{\bm{Z}}_{N}]^{-1}[\bm{\alpha}][\bm{Z}_{F2}]]^{-1}.
	\end{aligned}
\end{equation}
\begin{equation}
\label{eq:22}
\begin{split}
[\bm{I}+\bm{W}_3 ]^{-1}=[\bm{I}+[\bm{I}+\bm{V}_2 ]^{-1} [\bm{V}_3 ]]^{-1}\\
=[\bm{I}+[\bm{I}+[\bm{I}+\bm{U}_1 ]^{-1}[\bm{U}_2 ]]^{-1}[\bm{I}+\bm{U}_1 ]^{-1}[\bm{U}_3 ]]^{-1}\\
=[\bm{I}+[\bm{I}+[\bm{I}+ [\tilde{\bm{Z}}_{N}]^{-1}[\bm{\alpha}][\bm{Z}_{F1}]]^{-1}[\tilde{\bm{Z}}_{N}]^{-1} [\bm{\alpha}][\bm{Z}_{F2} ]]^{-1} \\
[\bm{I}+[[\tilde{\bm{Z}}_{N}]^{-1} [\bm{\alpha}][\bm{Z}_{F1}]]^{-1}[\tilde{\bm{Z}}_{N}]^{-1}[\bm{\bm{\alpha}}][\bm{Z}_{F3} ]]^{-1}.
\end{split}
\end{equation}
From equations (20), (21), and (22), it can be observed that the solution of these equations is dependent on that level and the lower levels of the binary tree block interaction matrix. At each level, the inverse of the matrix system equation can be efficiently computed by using a fast power series solution\cite{18}. The fast power series iterative solution converges in two fixed iterations. The  solution process only depends on the matrix-vector product of the H-Matrix, thus retaining the complexity of $O(NlogN)$\cite{18}. The ML solution can be computed at the desired level per the required accuracy. Our results show that the solution at the leaf level gives an accurate result leading to time and memory savings. 
\section{NUMERICAL RESULTS}
In this section, we show the accuracy and efficiency of the proposed method. The simulations are carried out on 128 GB memory and an Intel (Xeon E5-2670) processor system for the double-precision data type. The H-Matrix computation is done with the ACA matrix compression error tolerance of 1e-3 \cite{22} and solved with GMRES iterative solver with convergence tolerance of 1e-6 \cite{12}.  For a compressed or dense matrix $[\textbf{Z}]$ if we want to expand $[1+\textbf{Z}]^{-1}$ in power series, the necessary and sufficient condition for convergence is $|\textbf{Z}|<1$ and we choose 0.1 for our simulations  \cite{25}.The conductor and dielectric geometry with dielectric constant $\epsilon_r$ is meshed with an element size less than $\lambda/10$ and $\lambda/(10\sqrt{\epsilon_r})$ respectively. To show the accuracy of the proposed method, the RCS results are compared with full H-Matrix iterative solver\cite{12}.  In the further subsections, we demonstrate the far-field memory and computation time savings along with in solution time saving with our proposed ML power series solution with different examples.
\subsection{PEC square plate}
To show the accuracy and efficiency on a PEC object in this subsection, we consider a square plate of size 15.0 $\lambda$ along x and y axis meshed with 67,200 unknown edges. The square plate mesh is divided with binary tree division till level 6. The PEC S-EFIE H-Matrix  is solved with ML power series solution method and H-Matrix iterative solver. ML power series converges in 2 iterations, and the iterative solver solution converges in 686. Only the far-field matrix at leaf level 6 is computed for the ML power series solution, ignoring far-field computation from levels 1 to 5 of the binary tree.
\begin{figure}
  \begin{center}
  \includegraphics[width=8cm]{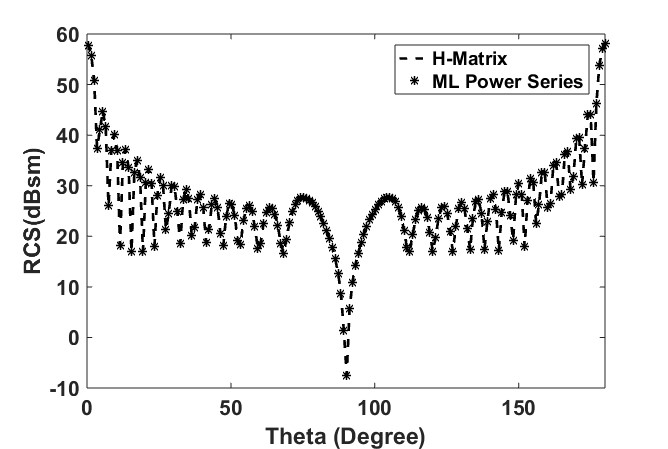}
  \caption{Bi-static RCS of the PEC square plate with VV polarized plane wave incident at $ \theta=0^\circ$, $\phi=0^\circ$, and observation angles $\theta=0^\circ$  to $180^\circ $, $\phi=0^\circ$.}
  \end{center}
\end{figure}
Fig. 2. shows the Bi-static RCS of a PEC square plate, and from the Fig., it can be observed that the solution with ML power series solver matches with the H-Matrix iterative solver. Table 1 shows the savings in memory, computation, and solution time of the ML power series solution method as compared with conventional H-Matrix-based iterative solver.
\begin{table}
\begin{center}
\caption{ Matrix memory, fill and solution time for a PEC square plate.}
\begin{tabularx}{\columnwidth}{X|X|X|X} \hline 
				&{\bf Memory (GB)} 	&{\bf Matrix Fill Time (H)} &{\bf Solution Time (sec)}\\  	 \hline
\multicolumn{1}{l|}{H-Matrix}	&\multicolumn{1}{c|}{5.04} &\multicolumn{1}{c|}{1.24} &\multicolumn{1}{c}{500.85}\\ 	\hline 
\multicolumn{1}{l|}{ML Power Series} &\multicolumn{1}{c|}{4.71} &\multicolumn{1}{c|}{1.08} &\multicolumn{1}{c}{3.95}\\ \hline
\end{tabularx}
\end{center}
\end{table}
\subsection{Dielectric slab}
To show the accuracy and efficiency for a considerable size dielectric problem in this subsection, we consider a dielectric slab elongated along the y-axis with a height of 10.0 $\lambda$ length, 1.0 $ \lambda $ width, and 0.1 $\lambda$ thickness and dielectric constant ($\epsilon_r=2.0$)  meshed with 120,080 tetrahedral faces. The ML power series converges in 2 iterations, and the regular H-Matrix iterative solver converges in 33 iterations.
\begin{figure}
  \begin{center}
  \includegraphics[width=8cm]{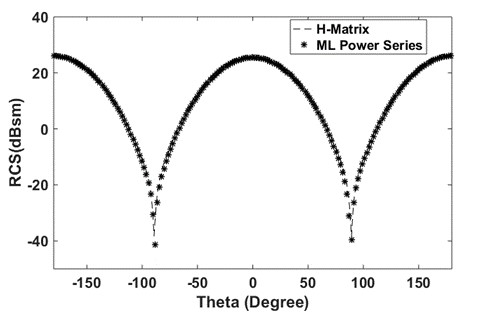}
  \caption{Bi-static RCS of the dielectric slab with VV polarized plane wave incident at $\theta=0^\circ$, $\phi=0^\circ$, and observation angles $\theta=-180^\circ$  to $180^\circ $, $\phi=0^\circ$.}
  \end{center}
\end{figure}
The dielectric slab mesh is divided with binary tree division till level 10. Only the far-field matrix at leaf level 10 is computed for the ML power series solution. The accuracy of the method for a  Bi-static RCS is shown in Fig. 3. Table 2 shows the significant matrix memory, matrix fill and solution time savings of the ML power series solution compared to the conventional H-Matrix-based iterative solver. 
\subsection{Dielectric hollow cylinder}
In this subsection, we consider a dielectric hollow cylinder elongated along the y-axis with a size of 6.0$\lambda$ length, 0.4$\lambda$ outer radii, and 0.05$\lambda$ thickness with a dielectric constant ($\epsilon_r=2.0$), meshed with 158,830 tetrahedral faces. The ML power series converges in 2 iterations, and the H-Matrix iterative solver converges in 24 iterations.\\
\begin{figure}
  \begin{center}
  \includegraphics[width=7cm]{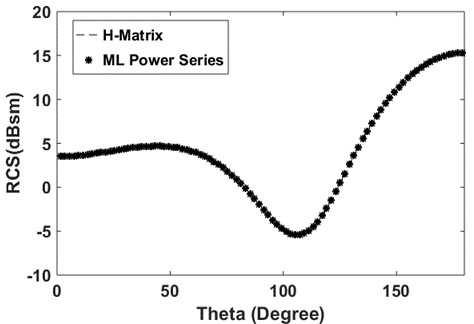}
  \caption{Bi-static RCS of a dielectric hollow cylinder with VV polarized plane wave incident at $\theta=0^\circ$, $\phi=0^\circ$, and observation angles $\theta=0^\circ$  to $180^\circ $, $\phi=0^\circ$. }
  \label{fig:figure_3}
  \end{center}
\end{figure}
\indent The hollow cylinder mesh is partitioned with a binary tree division till level 8, and for the ML power series solution only the far-field matrix at leaf level 8 is computed. Fig. 4. shows the close match in the bi-static RCS computed using the ML power series method and that with regular H-Matrix iterative solver. Table 3 shows the memory and time saving of the ML power series solution compared to the conventional H-Matrix iterative solver.
\section{CONCLUSION}
It can be observed from the illustrative examples in the previous sections that our proposed ML power series solution method gives considerable matrix memory, fill  and solve time saving for significant size problems. The solution method is as accurate as the H-Matrix  iterative solver. The savings may not be substantial for small-size mesh structures. Still, the method will give significant savings for large-size problems taken up for illustration and for complex and sizeable electrical problems like antenna arrays and complex composite structures. Also, the technique is entirely algebraic in nature and can apply to fast analytical solver-based methods like AIM and MLFMA. The matrix block in each level can be computed independently, and the solution of the method only depends on the matrix-vector product of the system matrix. Hence, the proposed method is amenable to efficient parallelization.

\begin{table}
\begin{center}
\caption{Matrix memory, fill and solution time for a dielectric slab.}
\begin{tabularx}{\columnwidth}{X|X|X|X} \hline 
				&{\bf Memory (GB)} 	&{\bf Matrix Fill Time (H)}	&{\bf Solution Time (sec)}\\	\hline
\multicolumn{1}{l|}{H-Matrix}	& \multicolumn{1}{c|}{2.09} &\multicolumn{1}{c|}{6.12} &\multicolumn{1}{c}{24.52}\\ \hline 
\multicolumn{1}{l|}{ML Power Series} 	&\multicolumn{1}{c|}{0.50} &\multicolumn{1}{c|}{1.46} &\multicolumn{1}{c}{7.50}\\ \hline
\end{tabularx}
\end{center}
\end{table}
\begin{table}
\begin{center}
\caption{Matrix memory, fill and solution time for a dielectric hollow cylinder.}
\begin{tabularx}{\columnwidth}{X|X|X|X} \hline 
				&{\bf Memory (GB)}	&{\bf Matrix Fill Time (H)} &{\bf Solution Time (sec)} \\	\hline
\multicolumn{1}{l|}{H-Matrix} &\multicolumn{1}{c|}{3 .38} &\multicolumn{1}{c|}{10.00} &\multicolumn{1}{c}{54.52}\\	 \hline 
\multicolumn{1}{l|}{ML Power Series} &\multicolumn{1}{c|}{0.44} &\multicolumn{1}{c|}{1.26} &\multicolumn{1}{c}{16.16}\\ 	\hline
\end{tabularx}
\end{center}
\end{table}
\bibliographystyle{ACESJournal}
\bibliography{library}
%
\begin{ACESbiography}{Yoginder Kumar Negi }{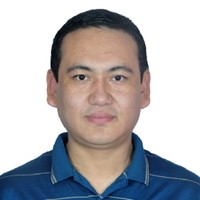}
obtained the B.Tech degree in Electronics and Communic-ation Engineering from Guru Gobind Singh Indraprastha University, New Delhi, India, in 2005, M.Tech degree in Microwave Electronics from Delhi University, New Delhi, India, in 2007 and the PhD degree in engineering from Indian Institute of Science (IISc), Bangalore, India, in 2018.

Dr Negi joined Supercomputer Education Research Center (SERC), IISc Bangalore in 2008 as a Scientific Officer. He is currently working as a Senior Scientific Officer in SERC IISc Bangalore. His current research interests include numerical electromagnetics, fast techniques for electromagnetic application, bio-electromagnetics, high-performance computing, and antenna design and analysis.
\end{ACESbiography}
\begin{ACESbiography}{B. Narayanaswamy}{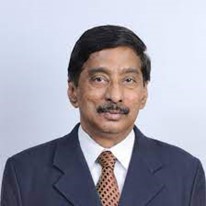}
received the B.E. degree (Hons.) in Electronics and Communi-cation from the University of Madras, Chennai, India, in 1972, and the Ph.D. degree from the Indian Institute of Science, Bengaluru, India, in 1979.

He joined the Department of Aerospace Engineering, Indian Institute of Science, as an Assistant Professor, in 1981, where he became a Full Professor in 1991, served as the Associate Director, from 2005 to 2014, and is currently an INSA Senior Scientist at the Supercomputer Education and Research Centre. He has authored over 200 publications in the international journals and international conferences. His current research interests include numerical electromagnetics, high-performance computing and networks, polarimetric radars and aerospace electronic systems, information security, and digital library.

Dr. Narayanaswamy is a fellow of the World Academy of Sciences (TWAS), the National Academy of Science, the Indian Academy of Sciences, the Indian National Academy of Engineering, the National Academy of Sciences, and the Institution of Electronics and Telecommunication Engineers.
\end{ACESbiography}
\begin{ACESbiography}{Sadasiva M. Rao }{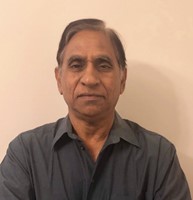}
obtained his Bachelors, Masters, and Doctoral degrees in electrical engineering from Osmania University, Hyderabad, India, Indian Institute of Science, Bangalore, India, and University of Mississippi, USA, in 1974, 1976, and 1980, respectively.  He is well known in the electromagnetic engineering community and included in the Thomson Scientifics \textit{Highly Cited Researchers List}. 

Dr. Rao has been teaching electromagnetic theory, communication systems, electrical circuits, and other related courses at the undergraduate and graduate level for the past 30 years at various institutions. At present, he is working at Naval Research Laboratories, USA. He published/presented over 200 papers in various journals/conferences. He is an elected Fellow of IEEE.
\end{ACESbiography}
%
\end{document}